\documentclass[aps,pre,twocolumn,showpacs]{revtex4}

\usepackage{amstext,amsopn,amsmath,amssymb}
\usepackage{epsfig}

\begin{document}

\title{Kinetics in one--dimensional lattice gas and Ising models
      from time--dependent \\density functional theory}

\author{M. Kessler, W. Dieterich,}
\affiliation{Fachbereich Physik Universit\"{a}t Konstanz, D-78457 Konstanz, Germany}

\author{H.L. Frisch,}
\affiliation{Dept. of Chemistry, SUNY at Albany, Albany NY 12222, USA}

\author{J.F. Gouyet}
\affiliation{Laboratoire de Physique de la Mati\`ere
Condens\'ee 
Ecole Polytechnique, 91128 Palaiseau Cedex, France}

\author{and P. Maass}
\affiliation{Fachbereich Physik Universit\"{a}t Konstanz, D-78457 Konstanz, Germany}

\date{January 23, 2002}

\begin{abstract}
Time--dependent density functional theory, proposed recently in the
context of atomic diffusion and non--equilibrium processes in solids,
is tested against Monte Carlo simulation. In order to assess the
basic approximation of that theory, the
representation of non--equilibrium states by a local
equilibrium distribution function, we focus on one--dimensional
lattice models, where all equilibrium properties can be worked exactly
from the known free energy as a functional of the density. This
functional determines the thermodynamic driving forces away from
equilibrium. In our studies of the interfacial kinetics of atomic
hopping and spin relaxation, we find excellent agreement with
simulations, suggesting that the method is useful also for
treating more complex problems.
\end{abstract}


\maketitle

\section{Introduction}

Establishing a link between macroscopic laws of diffusion and
relaxation with a microscopic master equation for atomic degrees of
freedom has remained a fundamental problem in non--equilibrium
statistical mechanics. Moreover, numerous examples exist in
metallurgy, semiconductor device technology, glass science and polymer science, where control over the
time--development of microstructures is crucial in the design of
materials with special mechanical, electrical and magnetic
properties. Hence there is a need also from a practical viewpoint to
derive tractable kinetic equations incorporating specific materials
properties, so that processes like nucleation, spinodal decomposition
and magnetic relaxation can be treated in a realistic manner.

The simplest approach is to study mean--field kinetic equations, derived
from the master equation by neglecting any atomic
correlation effects \cite{F91,Bi74,Pl97}. Such equations suffer from the obvious drawback
that their stationary solutions yield the mean--field phase diagram,
which may differ even qualitatively from the correct phase
diagram. Several routes for 
improvement have been proposed in the
literature, including the path probability method \cite{Ki79,Re94}, effective
Hamiltonian methods \cite{Na} or time--dependent density functional
theory (TDFT) \cite{Rei96,Fi98,Die90}. The latter approach implements the idea of local
equilibrium and leads to thermodynamic driving forces which in
principle are
derived from an exact free energy functional. Density functional
theories are normally formulated for continuous fluid systems \cite{Loe93}, but
adaptation to discrete lattice systems is straightforward \cite{Ni93,Rei94}. In
this way one can obtain generalized mean--field kinetic equations for
single--particle or single--spin densities which in principle are
consistent with the exact equilibrium properties. If necessary,
additional approximations with respect to equilibrium quantities can
be carried out in a separate step.\\
The above--mentioned kinetic theories mostly focus on purely
dissipative processes in discrete lattice systems. For fluid systems,
the derivation of nonlinear transport equations for hydrodynamic
variables is a more complicated subject. Some generalizations of
Moris' well--known projection operator technique and mode coupling
approximations to situations far away from equilibrium
\cite{KAW73,GRA82} have been applied, for example, to problems of nonlinear
hydrodynamics, the glass transition \cite{LAT00} and to granular flows \cite{SCHO94}.\\
Our aim in this paper is to apply the TDFT scheme to purely
dissipative ``conserved'' atomic
migration and ``non--conserved'' spin dynamics processes in
one--dimensional lattice models for which the exact free energy
functional is known. This enables us to separate out and to test the
local equilibrium assumption against Monte Carlo
simulation. Specifically, we study the temporal evolution of domains
with different ordering, and of the associated interface. It is
demonstrated that in these problems the TDFT shows remarkable
quantitative accuracy, suggesting that this method may be useful also
under more general conditions.

In section 2 we briefly recall our basic approach. Section 3 starts
out with an exact free energy functional for a one--dimensional
lattice and provides expressions for local correlators in terms of
particle densities. With these results we arrive at a closed system of
kinetic equations on the single--particle level. Following the
classification by Hohenberg and Halperin \cite{Ho77} these equations take
the form of generalized ``model B'' or ``model A'' equations in cases
of a conserved or a non--conserved order parameter, respectively. Solving these
equations, we subsequently discuss the time--evolution of an initially
sharp interface between two differently ordered domains in both of these cases
(section 4). Excellent agreement with Monte Carlo simulations is
found, in contrast to ordinary mean--field (MF) theory which produces
substantial deviations.
  
\section{Overview of time--dependent density functional theory
  (TDFT) for stochastic lattice systems}

\subsection{Atomic hopping}

Let us begin with hopping of particles on a lattice of equivalent
sites $i$, which are either simply occupied ($n_{i}=1$) or vacant
($n_{i}=0$), so that occupation numbers satisfy $n_{i}^{2}=n_{i}$. The
hopping process is described by a master equation for probabilities
$P({\bf n},t)$ of finding an occupational configuration ${\bf
  n}\equiv\{n_{i}\}$ at time $t$. As elementary steps we assume moves
of a single particle from an occupied site to a vacant nearest
neighbor site. The associated rates $w_{i,k}({\bf n})$ for adjacent
sites $i$ and $k$ to exchange their occupation satisfy the detailed
balance condition with respect to a given lattice gas Hamiltonian
$H({\bf n})$.

A detailed description of TDFT is found in Ref. \cite{Fi98}. Hence we need
to recall only the main steps, and add some remarks as to their physical
content. The basic approximation is to replace the
distribution $P({\bf n},t)$
by the local equilibrium distribution
\begin{equation}\label{eq:Ploc}
P^{loc}({\bf n},t)=\frac{1}{Z(t)}\exp\left\{-\beta\left[H({\bf
    n})+\sum_{i}h_{i}(t)n_{i}\right]\right\},
\end{equation}
where deviations from equilibrium are represented in
terms of time--dependent single--particle fields ${\bf
  h}(t)=\{h_{i}(t)\}$. $Z(t)$ is a normalization factor, which at
equilibrium (${\bf h}(t)=0$) reduces to the canonical partition
function. Requiring self--consistency on the single--particle level
allows us to eliminate ${\bf h}(t)$ in favor of mean occupation
numbers ${\bf p}(t)=\{p_{i}(t)\}$ with $p_{i}(t)=\langle
n_{i}\rangle_{t}$, where $\langle\ldots\rangle_{t}$ denotes an average
with respect to the distribution (\ref{eq:Ploc}). In this way a closed system
of equations for ${\bf p}(t)$ can be derived.

To carry through this program we start from the equation of continuity
which follows directly from the original master equation. Replacement
of exact averages by local equilibrium averages gives
\begin{equation}\label{eq:dpi}
\frac{dp_{i}(t)}{dt}+\sum_{k}\langle j_{i,k}\rangle_{t}=0,
\end{equation}
with known expressions \cite{Fi98} for the current $j_{i,k}({\bf n})$ from site $i$
to site $k$ in terms of $w_{i,k}({\bf n})$. Notice that at any instant
of time the exponent in (\ref{eq:Ploc}) describes an inhomogeneous lattice
gas which involves a spatially varying single--particle potential
${\bf h}(t)$. Hence, calculation of averages from (\ref{eq:Ploc}) is
precisely the kind of problem treated by density functional theory
(DFT) in classical statistical mechanics. There, one considers a class
of systems with fixed interactions and arbitrary single--particle
potentials, specified here by $H({\bf n})$ and ${\bf h}$,
respectively. Averaged occupation numbers and correlation functions
are determined from derivatives of a free energy functional
$F({\bf p})$ associated with the Hamiltonian $H({\bf
  n})$. Specifically, ${\bf p}(t)$ is determined by the set of
equations
\begin{equation}\label{eq:hit}
h_{i}(t)+\mu_{i}({\bf p}(t))=\mu_{tot},
\end{equation}
with $\mu_{tot}$ the overall chemical potential, and
\begin{equation}
\mu_{i}({\bf p})=\partial F({\bf p})/\partial p_{i},
\end{equation}
the local chemical potential as functional of ${\bf p}$. Much
experience has been gained during the last two decades how to construct
$F({\bf p})$ from a given Hamiltonian $H({\bf n})$. In the subsequent
considerations we therefore assume $F({\bf p})$ to be
known. Since in the framework of DFT occupational correlation
functions are functionals of ${\bf p}$, we can formally regard
(\ref{eq:dpi}) as the desired closed set of equations for ${\bf p}(t)$.

In order to make this procedure explicit and to establish a
connection with thermodynamic driving forces, we again recall
Ref. \cite{Fi98} where it is shown that the average current can be written
as
\begin{equation}\label{eq:jik}
\langle j_{i,k}\rangle_{t}=M_{i,k}(t)[A_{i}(t)-A_{k}(t)].
\end{equation}
The quantities
\begin{equation}\label{eq:Ait}
A_{i}(t)=\exp[\beta\mu_{i}({\bf p}(t))]
\end{equation}
are local activities, whose discrete gradient (along the bond
connecting $i$ and $k$) plays the role of a thermodynamic force
that drives the current. The quantity
\begin{equation}\label{eq:mik}
M_{i,k}(t)=\frac{1}{2}\left\langle w_{i,k}({\bf
    n})\exp\left[\beta(h_{i}(t)n_{i}+h_{k}(t)n_{k})\right]\right\rangle_{t},
\end{equation}
where $M_{i,k}(t)=M_{k,i}(t)$, is a mobility coefficient
that depends on the actual nonequilibrium state. Further discussion
of (\ref{eq:mik}) simplifies when we choose the hopping rates $w_{i,k}({\bf
  n})$ such that they depend only on the energy in the initial state,
i.\,e.
\begin{equation}\label{eq:wik}
w_{i,k}({\bf n})=\alpha[n_{i}(1-n_{k})e^{\beta
  H_{i}}+n_{k}(1-n_{i})e^{\beta H_{k}}].
\end{equation}
The first term describes hopping from $i$ to $k$, with a thermally
activated rate determined by the interaction energy $H_{i}$ of a
particle at site $i$ with its environment. $\alpha$ is some bare rate
constant. The reverse hopping process is described by the second term
in (\ref{eq:wik}). With this expression for $w_{i,k}({\bf n})$, one
can show \cite{Fi98} that Eq. (\ref{eq:mik}) transforms into 
\begin{equation}\label{eq:Mik}
M_{i,k}(t)=\alpha\langle(1-n_{i})(1-n_{k})\rangle_{t}.
\end{equation}
At this stage, $M_{i,k}(t)$ does no longer explicitly depend on ${\bf
  h}$. Physically, equation (\ref{eq:Mik}) tells us that the mobility
  coefficient based on (\ref{eq:wik}) is
  given by the nearest--neighbor vacancy
correlator.

It should be kept in mind that in this TDFT--scheme all deviations
from equilibrium are described in a mean--field manner in the sense that the
underlying distribution function (\ref{eq:Ploc}) deviates from the canonical
distribution merely by single--particle terms. Relationships between
occupational correlators and densities ${\bf p}(t)$, which enter this
theory, are local in time and are given by the equilibrium
theory. This implies the assumption that correlators relax fast to
their local equilibrium values, compared with time scales characterizing
the evolution of ${\bf p}(t)$. Ordinary
kinetic mean--field theory is recovered when we use mean--field
expressions for $\mu_{i}({\bf p})$ and replace (\ref{eq:Mik}) by
$M_{i,k}^{MF}(t)=\alpha(1-p_{i}(t))(1-p_{k}(t))$. By contrast, in TDFT
the local chemical potential appearing in (\ref{eq:Ait}) is defined by the exact chemical potential functional so that (\ref{eq:dpi}) together
with (\ref{eq:jik}) describes relaxation towards the exact equilibrium
state. Moreover, the expression (\ref{eq:Mik}) for the mobility preserves
local correlation effects in the jump dynamics.

\subsection{Spin relaxation}

To exemplify the dynamics of a non--conserved order
parameter, we study spin--relaxation in a kinetic Ising model. Elementary transitions in the underlying master equation
are supposed to be individual spin flips $\sigma_{i}\rightarrow
-\sigma_{i}$, where $\sigma_{i}=\pm 1$. By $w_{i}({\boldsymbol\sigma})$ we
denote the associated rate in an initial spin configuration
${\boldsymbol\sigma}$. The local equilibrium distribution
$P^{(loc)}({\boldsymbol\sigma},t)$ is analogous to (\ref{eq:Ploc}). It
involves the Ising Hamiltonian $H(\boldsymbol\sigma)$ supplemented by time--dependent magnetic fields
${\bf h}(t)$, which couple to the spins in the form
$-\sum_{i}h_{i}(t)\sigma_{i}$. As shown in the Appendix, the equations
of motion read
\begin{equation}\label{eq:dsi}
\frac{d\langle\sigma_{i}\rangle_{t}}{dt}=-\Gamma_{i}(t)\sinh\beta\left(\frac{\partial
  F(\langle{\boldsymbol\sigma}\rangle_{t})}{\partial\langle\sigma_{i}\rangle_{t}}-h\right),
\end{equation}
with kinetic coefficients
\begin{equation}\label{eq:Gamma}
\Gamma_{i}(t)=2\langle w_{i}({\boldsymbol\sigma})e^{-\beta
  h_{i}(t)\sigma_{i}}\rangle_{t},
\end{equation}
$F$ is the intrinsic free energy functional associated with the
exchange interaction, and $h$ an overall constant magnetic
field. Equation (\ref{eq:dsi}) again displays the exact thermodynamic driving force in the
spirit of TDFT. It can be regarded as a generalized ``model A'' equation
in the classification by Hohenberg and Halperin \cite{Ho77}, whereas
Eqs. (\ref{eq:dpi}), (\ref{eq:jik}) and (\ref{eq:Ait}) constitute
generalized ``model B'' equations. Note that sufficiently close to
equilibrium one can ignore ${\bf h}(t)$ in (\ref{eq:Gamma}) and
linearize the sinh--term in (\ref{eq:dsi}) to obtain
$d\langle\sigma_{i}\rangle_{t}/dt\simeq -2\beta\langle
w_{i}(\boldsymbol\sigma)\rangle_{eq}(\partial
F/\partial\langle\sigma_{i}\rangle_{t}-h)$. The kinetic coefficient is
then simply given by the equilibrium spin flip rate $\langle
w_{i}(\boldsymbol{\sigma})\rangle_{eq}$.

\subsection{Consistency with thermodynamics}

Finally it is easy to show that our evolution equations are
consistent with the second law of thermodynamics: The total free
energy decreases monotonously with time until the equilibrium
condition is satisfied. For hopping, the rate of change of the free
energy is given by
\begin{eqnarray}\label{eq:dF}
\frac{dF}{dt} \hspace{-0.2cm}& = \hspace{-0.2cm}& \sum_{i}\frac{\partial F}{\partial
  p_{i}}\frac{dp_{i}}{dt}\nonumber\\
\hspace{-0.2cm}& =\hspace{-0.2cm} &
  -k_{B}T\sum_{i,k}M_{i,k}x_{i}(e^{x_{i}}-e^{x_{k}})\nonumber\\
\hspace{-0.2cm}& = \hspace{-0.2cm}&
  -\frac{k_{B}T}{2}\sum_{i,k}M_{i,k}(x_{i}-x_{k})(e^{x_{i}}-e^{x_{k}})\le
  0,
\end{eqnarray}
where we have used (\ref{eq:dpi}), (\ref{eq:jik}) and (\ref{eq:Ait}) together
with the abbreviation $\partial(\beta F/\partial p_{i})=x_{i}$, and
$M_{i,k}=M_{k,i}$. Currents through the boundaries of the system are
supposed to be zero. The inequality in (\ref{eq:dF}) arises from
$M_{i,k}>0$, see (\ref{eq:mik}), and from $(x-y)(e^{x}-e^{y})>0$ for $x\ne
y$. The equality sign in (\ref{eq:dF}) holds if and only if
$x_{i}=x_{k}$ for all $i$ and $k$, which means that $\mu_{i}=\rm
const$.

Similarly, for the kinetic Ising spin model, the total free
energy including the coupling to the external field $h$ satisfies
\begin{equation}\label{eq:ddt}
\frac{d}{dt}\left(F-h\sum_{i}\sigma_{i}\right)=-k_{B}T\sum_{k}\Gamma_{k}x_{k}\sinh
x_{k}\le 0,
\end{equation}
where $x_{k}=\beta(\partial
F/\partial\langle\sigma_{k}\rangle_{t}-h)$. The inequality follows because
$\Gamma_{k}>0$ (see (\ref{eq:Gamma})) and $x\sinh x>0$ for $x\ne
0$. Equation (\ref{eq:ddt}) becomes an equality if $\partial
F/\partial\langle\sigma_{k}\rangle_{t}=h$ for all $k$.

\section{One dimension: exact functionals}

To test the local equilibrium
distribution (\ref{eq:Ploc}) it is desirable to avoid any
approximation with respect to static properties. This can be achieved
by using exact free energy functionals, which are available
for certain one--dimensional systems \cite{Per,Bu1,Bu2}.

\subsection{Atomic hopping}

For a lattice gas with nearest neighbor interactions on a linear
chain of sites $i$; $1\le i\le M$; with
occupied boundary sites at $i=0$ and $i=M+1$, the free energy functional reads \cite{Bu1}
\begin{eqnarray}\label{eq:Fp}
F\{{\bf p}\}\hspace{-0.2cm}&=&\hspace{-0.2cm}V\sum_{i=0}^{M}p_{i+1,i}^{(1)}
+k_{B}T\sum_{i=0}^{M-1}\left[\sum_{n=1}^{4}p_{i+1,i}^{(n)}\ln
  p_{i+1,i}^{(n)}-p_{i}\ln p_{i}\right. \nonumber\\
&& \hspace{-0.8cm} \left. \phantom{\sum_{n=1}^{4}} -(1-p_{i})\ln(1-p_{i})\right],
\end{eqnarray}
where $V$ denotes the interaction constant, and
$p_{i+1,i}^{(n)}$ with $n=1,\ldots 4$ are the two--point
correlators for the four possibilities of particles and holes on site
$i$ and site $i+1$,
\begin{eqnarray}\label{eq:pi+1}
p_{i+1,i}^{(1)} \hspace{-0.2cm}& = \hspace{-0.2cm}& \langle n_{i+1}n_{i}\rangle, \nonumber\\
p_{i+1,i}^{(2)} \hspace{-0.2cm}& = \hspace{-0.2cm}& \langle(1-n_{i+1})n_{i}\rangle=p_{i}-p_{i+1,i}^{(1)}, \nonumber\\
p_{i+1,i}^{(3)} \hspace{-0.2cm}& = \hspace{-0.2cm}& \langle n_{i+1}(1-n_{i})\rangle=p_{i+1}-p_{i+1,i}^{(1)}, \nonumber\\
p_{i+1,i}^{(4)} \hspace{-0.2cm}& = &\hspace{-0.2cm} \langle(1-n_{i+1})(1-n_{i})\rangle=1-p_{i}-p_{i+1}+p_{i+1,i}^{(1)}. \nonumber\\
\end{eqnarray}
From the techniques of Ref. \cite{Bu1} it follows that
\begin{equation}\label{eq:pi1}
p_{i+1,i}^{(1)}\;p_{i+1,i}^{(4)}=p_{i+1,i}^{(2)}\;p_{i+1,i}^{(3)}\;e^{-\beta
  V},
\end{equation}
a relation, which is equivalent to the quasi--chemical
approach. Given these relations, it turns out that $\partial F/\partial
p_{i+1,i}^{(1)}=0$. This suggests that $F$ may be minimized also with
respect to correlators in cases where these cannot be calculated explicitly.

The boundary conditions for the correlators are
$p_{1,0}^{(1)}=p_{1}$ and $p_{M+1,M}^{(1)}=p_{M}$. For $1\le i\le M-1$,
combination of
(\ref{eq:pi+1}) with (\ref{eq:pi1}) yields a quadratic equation for
$p_{i+1,i}^{(1)}$. In this way the representation of $p_{i+1,i}^{(n)}$
as functionals of ${\bf p}$ is completed. The fact that
$p_{i+1,i}^{(n)}$ only depends on $p_{i+1}$ and $p_{i}$ clearly is a
special feature in one dimension.

The kinetic equations derived in section 2 are now combined with
(\ref{eq:Fp}). Evidently, from (\ref{eq:Mik}),
\begin{equation}\label{eq:Mii}
M_{i,i+1}(t)=\alpha\, p_{i+1,i}^{(4)},
\end{equation}
while the local chemical potential is found to satisfy
\begin{equation}\label{eq:beta}
\beta\mu_{i}=\ln\frac{p_{i+1,i}^{(2)}\,p_{i,i-1}^{(3)}}{p_{i+1,i}^{(4)}\,p_{i,i-1}^{(4)}}-\ln\frac{p_{i}}{1-p_{i}}.
\end{equation}
From (\ref{eq:jik}), (\ref{eq:Mii}) and (\ref{eq:beta}) we obtain for the current 
\begin{eqnarray}\label{eq:jii}
\langle j_{i,i+1}\rangle_{t}&=&\alpha\left[\frac{p_{i,i-1}^{(3)}}{p_{i}}\frac{(1-p_{i})}{p_{i,i-1}^{(4)}}p_{i+1,i}^{(2)} \right. \nonumber\\
&& \left. -\frac{p_{i+2,i+1}^{(2)}}{p_{i+1}}\frac{(1-p_{i+1})}{p_{i+2,i+1}^{(4)}}p_{i+1,i}^{(3)}\right].
\end{eqnarray}
In these last equations (\ref{eq:Mii})--(\ref{eq:jii}), densities and correlators
are local equilibrium quantities. The final form of our kinetic
equations as a nonlinear set of differential equations for $p_{i}(t)$ emerges when we reexpress $p_{i+1,i}^{(n)}$ in terms of
$p_{i+1}$ and $p_{i}$ in the way described above.

For comparison we also consider the ordinary mean--field equations. These
are obtained by factorizing all correlators in (\ref{eq:beta}) and (\ref{eq:jii}), for example
$p_{i+1,i}^{(1)}\simeq p_{i+1}p_{i}$. The mean--field current is then
found as 
\begin{eqnarray}\label{eq:jMF}
\langle j_{i,i+1}^{MF}\rangle_{t}&=&\alpha\left\{p_{i}-p_{i+1}+K[p_{i-1}p_{i}(1-p_{i+1})\right.\nonumber\\
&&\left. -(1-p_{i})p_{i+1}p_{i+2}]\right\}
\end{eqnarray}
with $K=\exp(\beta V)-1$.

\subsection{Spin relaxation}

Next we turn to spin relaxation in the linear Ising model. Rather than using (\ref{eq:dsi}) we
immediately choose transition rates
\begin{equation}
w_{i}({\boldsymbol\sigma})=\frac{\alpha}{2}\left(1-\frac{\gamma}{2}\sigma_{i}(\sigma_{i+1}+\sigma_{i-1})\right)(1-\delta\sigma_{i})
\end{equation}
and start  from the evolution equations for single spins as given in the original
work by Glauber \cite{Gla},
\begin{eqnarray}
  \frac{d\langle\sigma_{i}\rangle_{t}}{dt} &=&
  \alpha \left[ \langle\sigma_{i}\rangle_{t} -
  \frac{\gamma}{2}(\langle\sigma_{i+1}
  \rangle_{t}+\langle\sigma_{i-1}\rangle_{t}) - \delta \right.  \nonumber\\
&& \left. +\frac{\delta\gamma}{2}(\langle\sigma_{i+1}\sigma_{i}\rangle_{t} +
  \langle\sigma_{i}\sigma_{i-1}\rangle_{t})\right].
\label{eq:dsit}
\end{eqnarray}
Here, $\gamma=\tanh 2\beta J$ and $\delta=\tanh\beta h$, where $J>0$
denotes the (ferromagnetic) nearest--neighbor exchange coupling and $h$ a constant
external magnetic field. It is well known that for $h=0$ these equations become linear and
easily soluble. By contrast, for $h\ne 0$, the appearance of correlators
$\langle\sigma_{i+1}\sigma_{i}\rangle_{t}$ in (\ref{eq:dsit}) prevents
us from obtaining an
exact solution. Using the
well--known representation of Ising spin variables by occupation
numbers, $\sigma_{i}=2n_{i}-1$, and vice versa, we can treat the
correlators $\langle\sigma_{i+1}\sigma_{i}\rangle$ in perfect analogy
to $\langle n_{i+1}n_{i}\rangle$. In particular, Eq. (\ref{eq:pi1}) with $J=4V$
transforms into a quadratic equation for
$\langle\sigma_{i\pm 1}\sigma_{i}\rangle$, whose solution, expressed in terms of
$\langle\sigma_{i\pm 1}\rangle$ and $\langle\sigma_{i}\rangle$, is
substituted into (\ref{eq:dsit}). This yields our TDFT--equation of motion
for spins. Likewise, we obtain from (\ref{eq:pi1}) the free energy as
functional of the spin density, which could be used in (\ref{eq:dsi}).

\section{Application to interfacial kinetics}

We now apply the TDFT to problems of the time--evolution of an
initially sharp interface between differently ordered domains on a
linear chain. Our
purpose is to present a quantitative comparison with both Monte Carlo
simulation and simple MF--theory with respect to density
profiles, spin--density profiles and the respective correlators.

\subsection{Atomic hopping}

\begin{figure}
\begin{center}
\epsfig{file=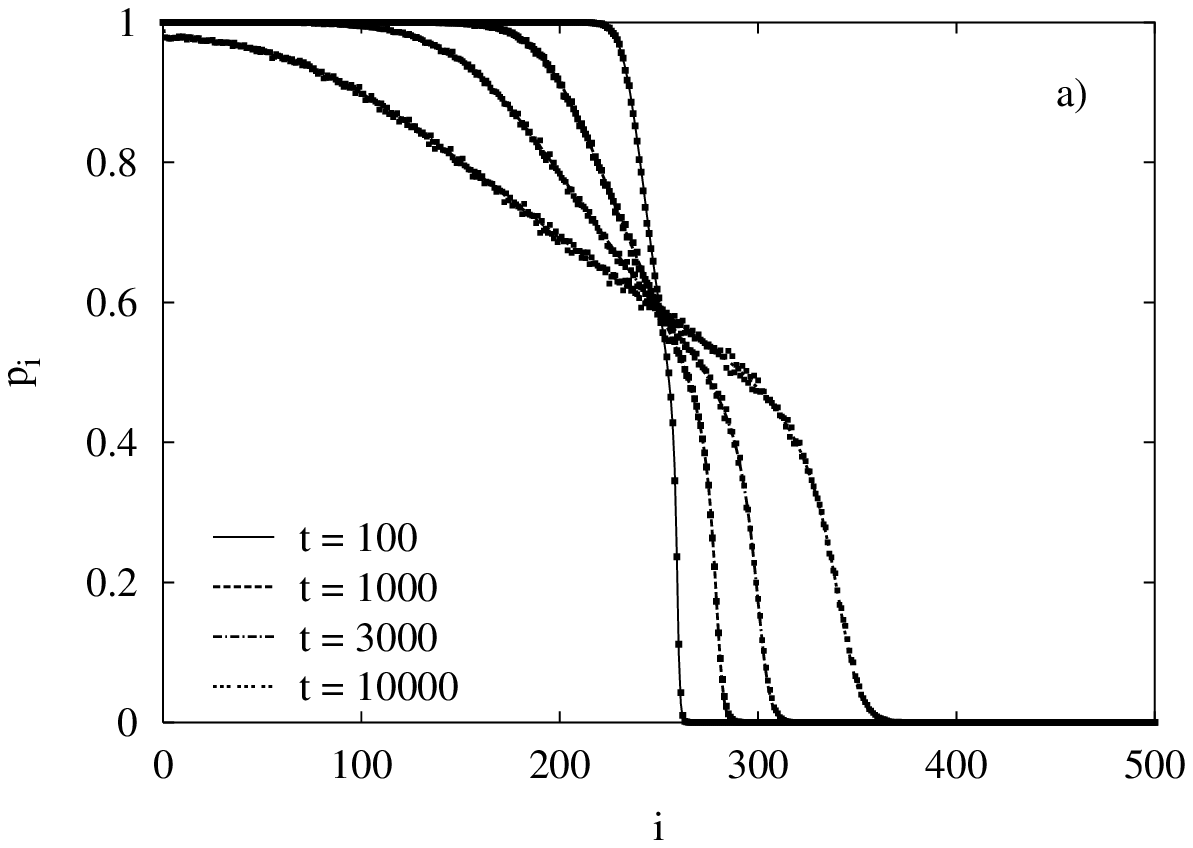,width=8cm}
\epsfig{file=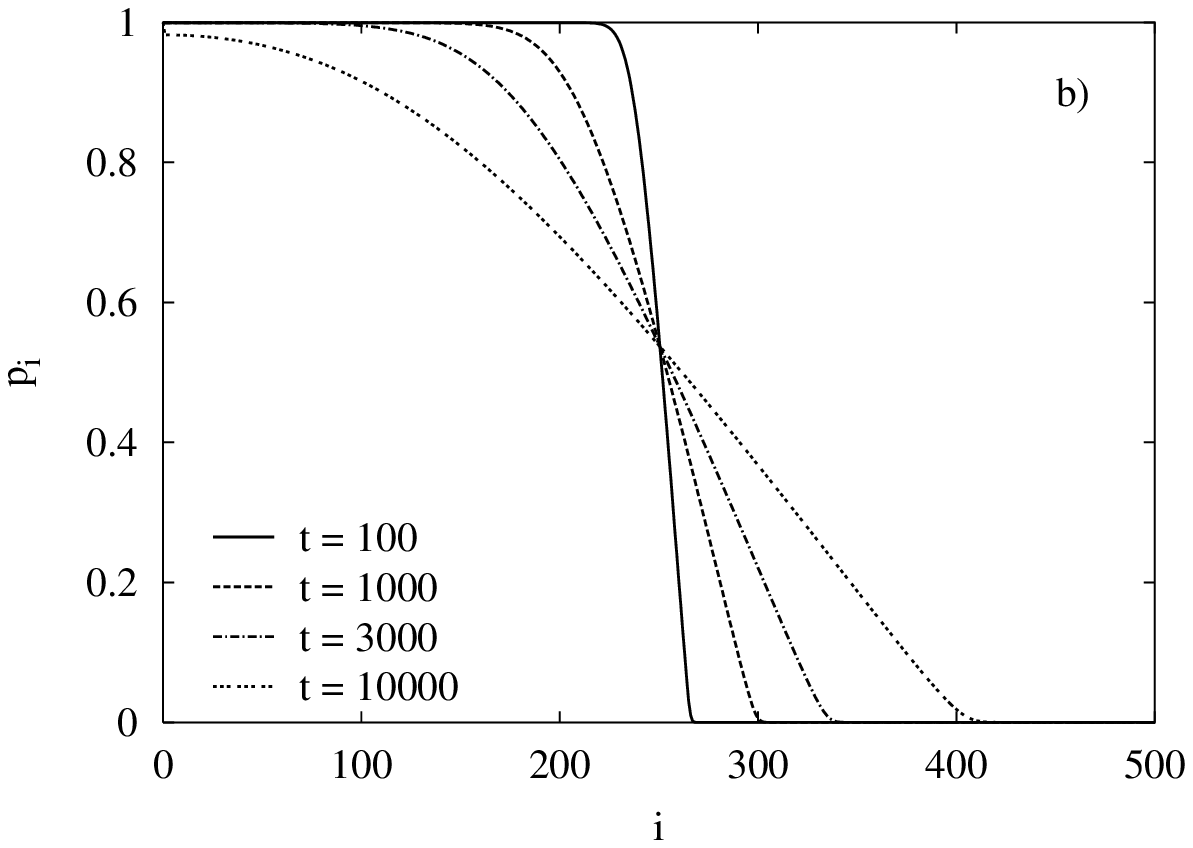,width=8cm}
\end{center}
\caption{Comparison of time--dependent density profiles $p_{i}$
in the case of hopping dynamics with repulsive interaction $\beta
V=3$, obtained from different methods.
a) TDFT (lines) and MC simulation (data points)
b) Kinetic MF--theory.}
\label{fig1}
\end{figure}

The length of
the chain is taken as $M=10^{3}$. As mentioned before, boundary sites
have fixed occupation $p_{0}=p_{M+1}=1$. Symmetrical initial conditions
at $t=0$ are chosen such that we have a vacant region centered around
the midpoint of the system, $p_{i}(0)=0$ for $250<i<750$, and complete occupation in
the complementary space. For $t>0$, the initially sharp density
profile will progressively broaden due to diffusion. This is shown in
Fig.~\ref{fig1} for $0\le i\le 500$ in the case of a repulsive interaction with $\beta
V=3$. Generally, the shape of profiles depends on how the interaction
enters the elementary hopping rates. Our choice (\ref{eq:wik}) implies that in
regions with densities $p\gtrsim 0.5$ a particle next to a vacant
target site has a large chance to be repelled by another particle and
hence will assume a large jump rate. By contrast, the repulsion will
be less active in dilute regions. This explains the asymmetry of the
profiles in Fig.~\ref{fig1}a, with a steep drop towards the empty region. The main
conclusion from Fig.~\ref{fig1}a is the perfect agreement between profiles from
TDFT, shown by the full lines, and from Monte Carlo (MC) simulation (data
points) \cite{Fus}. To get smooth profiles from simulation, we took averages
over $10^{4}$ Monte Carlo runs. By contrast, the MF--profiles in
Fig.~\ref{fig1}b are more symmetric and deviate significantly from those in
Fig.~\ref{fig1}a.

\begin{figure}
\begin{center}
\epsfig{file=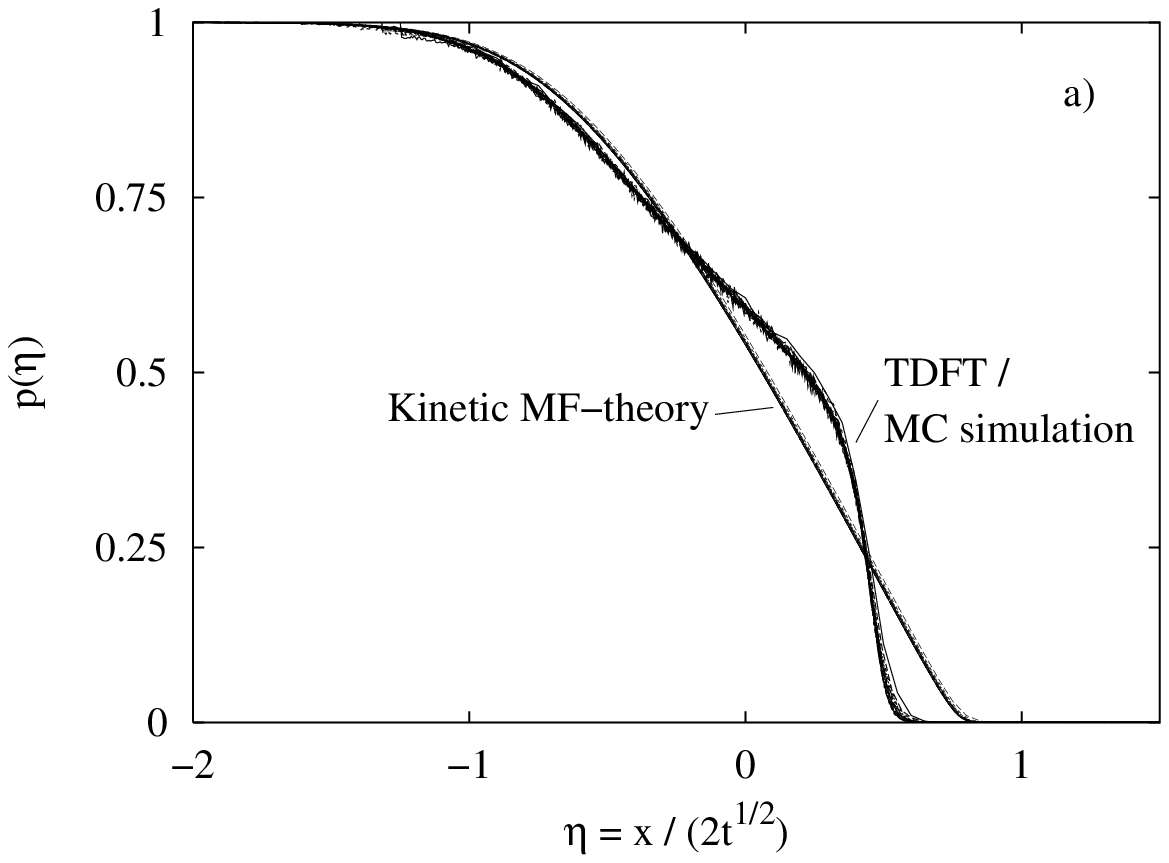,width=8cm}
\epsfig{file=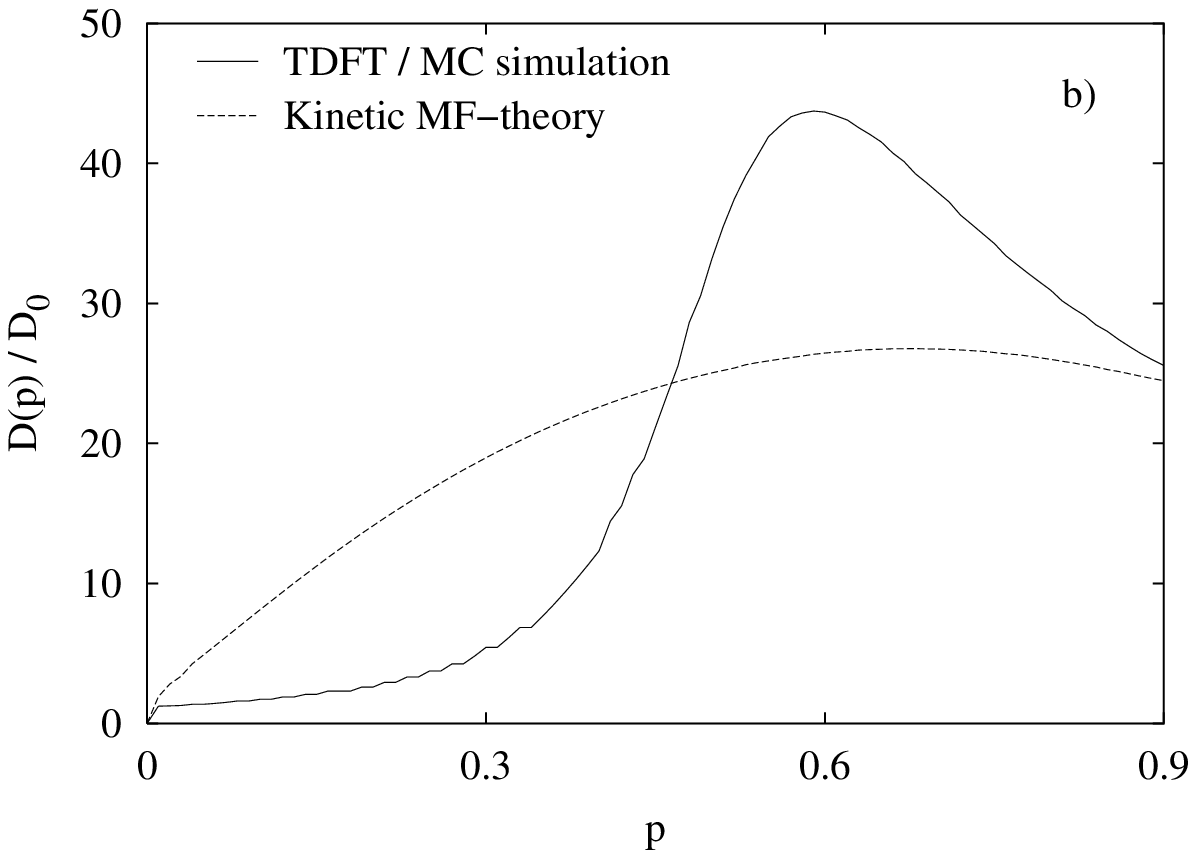,width=8cm}
\end{center}
\caption{a) Density profiles shown in Fig. 1 for different times against the scaling variable $\eta=x/(2\sqrt{t})$. The length $x$ in units of the lattice constant and $t$ in units of $\Delta t$, see footnote \cite{Fus}. b) Concentration--dependent diffusion coefficients $D(p)$ extracted from the master--curves of Fig. 2a by the Boltzmann--Matano method. (The normalization factor $D_{0}$ is the single--particle diffusion constant for infinite dilution.)}
\label{fig2}
\end{figure}

For diffusion processes on (continuous) length scales $x$ and time scales $t$
much larger than the elementary hopping distance and residence time, we
expect the density to depend only on the scaling variable
$\eta=x/(2\sqrt{t})$, provided the initial conditions can be expressed
in terms of $\eta$. This is verified in Fig.~\ref{fig2}a which shows
master--curves $p(\eta)$ obtained from the profiles in Fig.~\ref{fig1} for
different times. In this analysis the origin of the $x$--axis is
chosen to coincide with the initial density drop at $i=250$. As expected from
Fig.~\ref{fig1}, the TDFT master--curve, in contrast to MF--theory, practically
coincides with the Monte Carlo master--curve.

These results can be
analyzed further by the Boltzmann--Matano method \cite{Cr}, which
assumes a diffusion equation of the form $\partial p/\partial
x=\partial/\partial x(D(p(x))\partial p/\partial x)$ to hold. From the
master--curve $p(\eta)$ the concentration--dependent diffusion
coefficient $D(p)$ can be deduced according to 
\begin{equation}
D(p)=-\frac{2}{(dp/d\eta)}\int_{0}^{p}\eta(p')\,dp',
\end{equation}
where $\eta(p)$ is the inverse function of $p(\eta)$. The integral can be calculated accurately from the profile of Fig.~\ref{fig2}a
up to $p\simeq 0.9$, and the results for $D(p)$ are shown in
Fig.~\ref{fig2}b. Using the MF profile, we recover the $p$--dependence of the
mean--field diffusion constant. This quantity is calculated easily by
separating from the current (\ref{eq:jMF}) a factor $p_{i+1}-p_{i}$,
i.\,e. a discrete gradient of the density, and identifying the result
with Ficks' law. One obtains
\begin{equation}\label{eq:DMFp}
D^{MF}(p)=D_{0}(1+K[p^{2}+4p(1-p)]),
\end{equation}
with $D_{0}=\alpha a^{2}$, $a$ being the lattice spacing. The expression
(\ref{eq:DMFp})
shows a broad maximum around $p=2/3$, which reflects the average
effect of the repulsion of particles. The TDFT--diffusion constant,
however, shows a much sharper maximum. Moreover, when $p$ becomes
small, it approaches the value $D_{0}$ more rapidly, and thus gives
rise to the steepening of the density profile in the regime $p\lesssim\, 0.4$,
as observed in Fig.~\ref{fig2}a. This is a correlation
effect: In a dilute system, a fast hop of a particle due to the repulsion by a
neighboring particle is a rare event because nearest neighbor pairs get suppressed, $\langle
n_{i+1}n_{i}\rangle<p_{i+1}p_{i}$, and hence diffusion is slowed down
relative to the MF--prediction. This argument is supplementary to our previous discussion
of Fig.~\ref{fig1}a.

\begin{figure}
\begin{center}
\epsfig{file=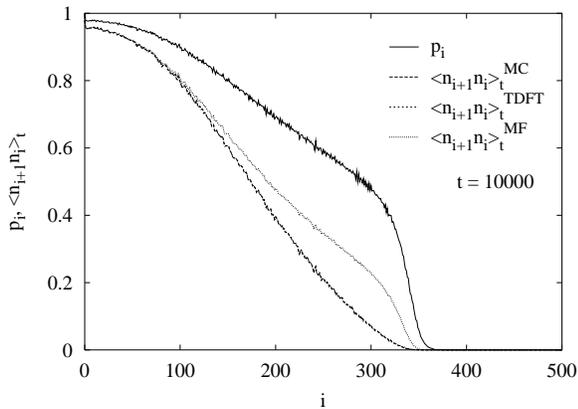,width=8cm}
\end{center}
\caption{Correlators $\langle n_{i+1}n_{i}\rangle_{t}$ at
$t=10^{4}$ from MC--simulation compared to correlators computed
from TDFT and MF--theory, using the same Monte Carlo density profile
$p_{i}$ (upper curve) as input. The TDFT--correlators are indistinguishable from
MC--correlators in this plot.}
\label{fig3}
\end{figure}

At this point we remark that such correlations induced by the
repulsion of particles are taken into account to a certain extent even
by MF--theory when applied to a two--sublattice structure. Density
profiles and effective diffusion coefficients calculated in this way
in a previous study \cite{Nas98} indeed are similar to those of the
present TDFT calculation shown in Fig.~\ref{fig2}a and \ref{fig2}b, respectively.

Because of the important role played by the correlators in TDFT it is
of interest to make a direct comparison with correlators obtained from
Monte Carlo simulation. Fig.~\ref{fig3} exemplifies perfect agreement between
those of TDFT and simulation, whereas MF--correlators, calculated here as
product $p_{i+1}p_{i}$ of the simulated densities, are significantly
larger when the densities are small.

\subsection{Spin relaxation}

\begin{figure}
\begin{center}
\epsfig{file=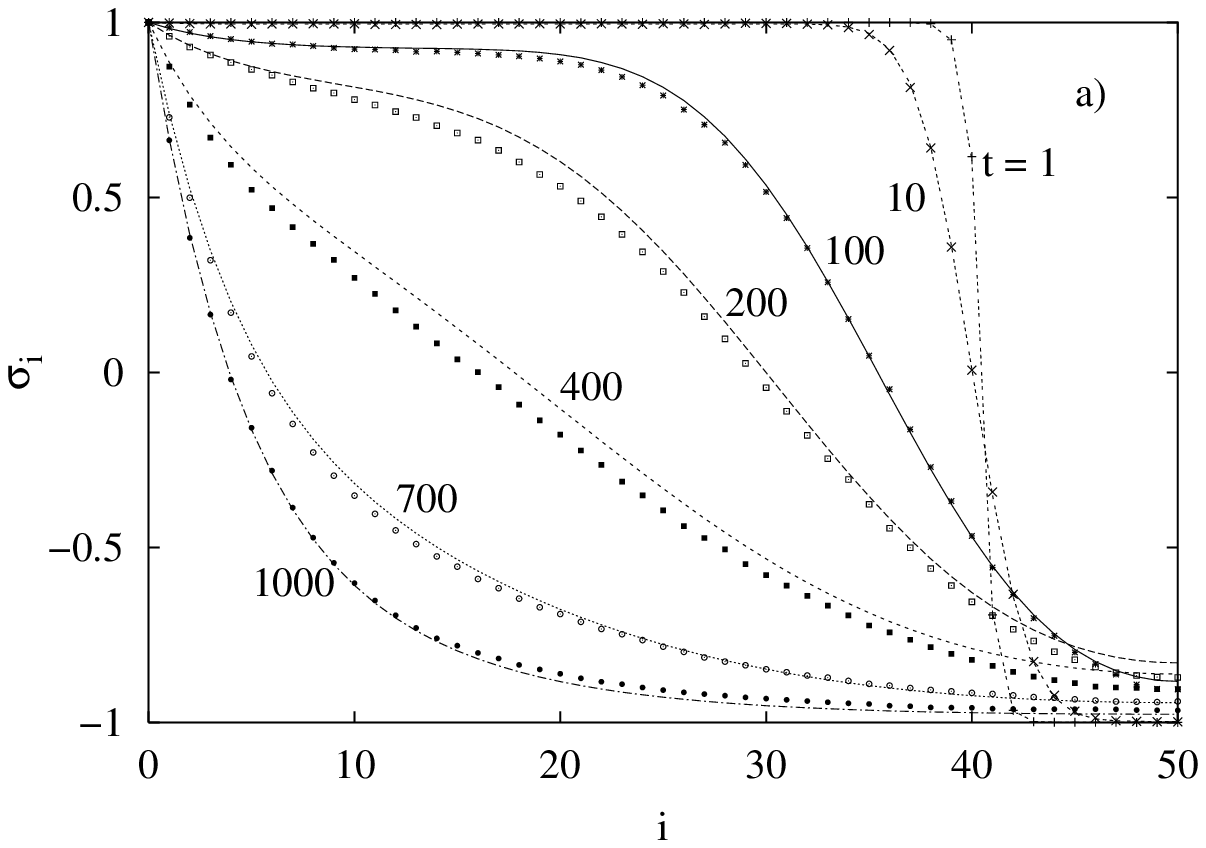,width=8cm}
\epsfig{file=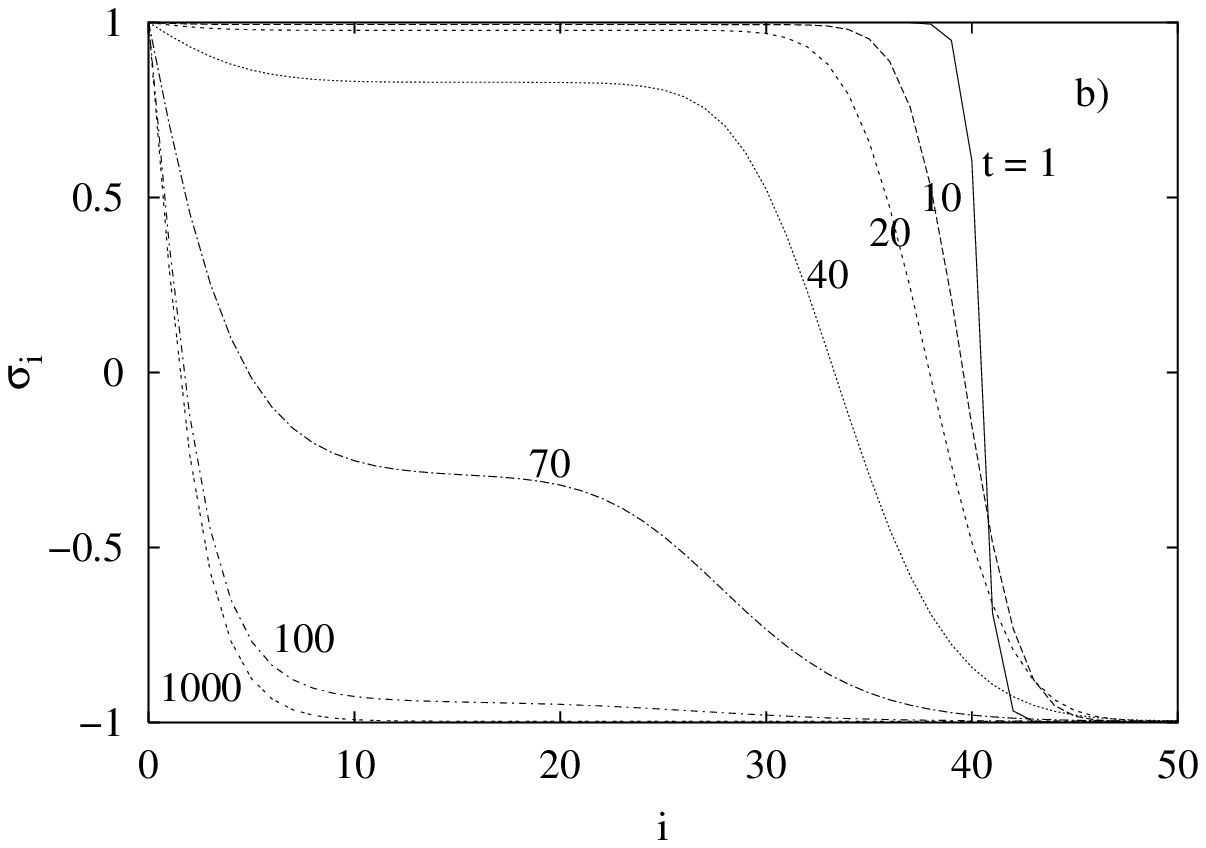,width=8cm}
\end{center}
\caption{Comparison of time--dependent spin density profiles in
the Glauber model with $\beta J=2$ and $\beta h=-0.1$, obtained from
different methods.
a) TDFT (lines) and MC--simulation (data points)
b) Kinetic MF--theory}
\label{fig4}
\end{figure}
In our study of ``non--conserved'' dynamics in the one--dimensional Glauber model \cite{Gla} we
choose a chain of length $M=10^{2}$ and fixed upward spins at the
boundaries, $\sigma_{0}=\sigma_{M+1}=1$. Our initial condition at
$t=0$ now is $\sigma_{i}=-1$ for $40\le i\le 60$ and $\sigma_{i}=+1$
for the remaining spins. Notice that in the case $h=0$ simple
MF--theory in the spirit of this work becomes exact because the
correlators in (\ref{eq:dsit}) drop out. To depart from this trivial
situation we introduce a small field with $\beta h=-0.1$ which favors downward
spin orientation. Spin density profiles in the region $0\le i\le 50$ for $\beta J=2$ at different
times $t>0$ are presented in Fig.~\ref{fig4}a, where the full lines correspond
to TDFT, and data points to simulation. The agreement is very good,
although not perfect. Generally, the spins in the interior of the
system relax towards the equilibrium in the external field. Spins near
the boundary are expected to relax towards an equilibrium profile which decays from
the boundary ($\sigma_{0}=1$) towards the interior ($\sigma_{i}\approx
-1$) on a length given by the correlation length $\xi$. For the
temperature considered, $\xi\simeq 5\,a$. 
\begin{figure}
\begin{center}
\epsfig{file=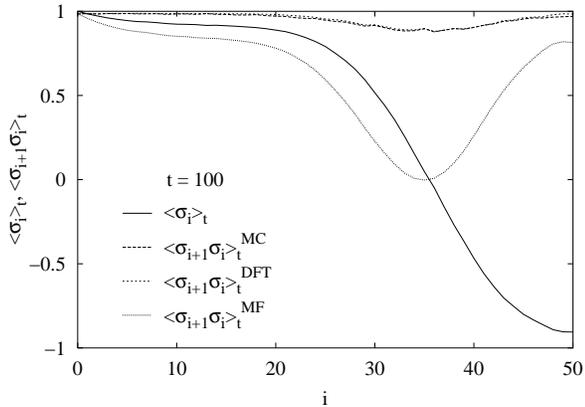,width=8cm}
\end{center}
\caption{Correlators $\langle\sigma_{i+1}\sigma_{i}\rangle_{t}$
at $t=10^{2}$ from MC--simulation compared to correlators computed
from TDFT and MF--theory, using the same MC spin density profile
$\langle\sigma_{i}\rangle_{t}$ as input. The TDFT--correlators are
nearly indistinguishable from MC--correlators in this plot.}
\label{fig5}
\end{figure}
During
the course of this relaxation, simple MF--theory, based on a
factorization of the last two terms in (\ref{eq:dsit}), gives quite
different results (Fig.~\ref{fig4}b). First of all, the overall process is much
faster than in Fig.~\ref{fig4}a. Second, within the region of the initial upspin domain it
predicts a constantly decreasing plateau which is not observed in
Fig.~\ref{fig4}a. The origin of these failures of MF--theory becomes clear when we
look at Fig.~\ref{fig5}: Monte Carlo-- and the almost identical TDFT--correlators
$\langle\sigma_{i+1}\sigma_{i}\rangle_{t}$ stay close to unity
throughout the system, in contrast to the MF--factorization, and
stabilize the respective spin configuration. Hence the relaxation
process progresses only by a successive broadening of the
interfacial region between the upspin and downspin domains and not by
a decaying plateau.

\section{Summary and outlook}

Applying a local equilibrium approximation to the master equation for
atomic or spin configurations, kinetic equations for particle or spin
densities were derived, which are compatible with the exact
thermodynamics. The derivation was largely based on concepts from
density functional theory. Kinetic equations obtained have the
form of generalized ``model B'' or ``model A'' equations in the
language of Ref. \cite{Ho77}, where thermodynamic driving forces originate
from the exact free energy functional. This ``time--dependent density
functional'' (TDFT)--scheme is tested against Monte Carlo simulations for
both a one--dimensional hopping model and the Glauber model, where the
exact free--energy functional is known. Studying the dynamics of the
interface between different domains, the TDFT yields excellent
agreement with simulations with respect to density or spin density
profiles and local correlation functions. The success of this theory
appears to be a consequence of the fast relaxation of correlators
towards their local equilibrium values.

Under the ultimate aim to develop theoretical tools for a description
of phase transformation processes in real materials, several
extensions of the present work are necessary. First of all, one needs
reliable approximations for the free energy functional in higher
dimensions. For two--dimensional lattice systems, a step in this
direction has been taken recently \cite{Bu3}, which was based on an extension of the techniques in Ref. \cite{Bu1}. Secondly, one would like to treat
multicomponent systems. In that case, local equilibrium distributions
of the type (\ref{eq:Ploc}) may be insufficient to describe interdiffusion
currents related to non--diagonal Onsager coefficients \cite{Na}. To
incorporate such effects into the TDFT--scheme is an open question
which deserves further study.

\section*{Acknowledgment}
The authors are grateful to J. Buschle for helpful
conversations. H.L.F. was supported by NSF Grant DMR 9628224 and
the Humboldt foundation. This work was supported in part by the Deutsche Forschungsgemeinschaft
(Ma 1636/2-1, SFB~513).

\section*{Appendix: Derivation of generalized ``model A''--equations}

The derivation of Eq. (\ref{eq:dsi}) proceeds in steps with some
similarity to Ref. \cite{Fi98}. In the present ``non--conserved'' case
we start from the master equation for single spin flips, see
e.\,g. Ref. \cite{Gla}, to obtain the time derivative of single--spin averages. Exact averages are in turn approximated by
averages $\langle\ldots\rangle_{t}$ with respect to the local
equilibrium distribution $P^{(loc)}(\boldsymbol\sigma,t)$, which has
the same form as (\ref{eq:Ploc}) apart from a sign change in the
second term in the exponent. (This is because the auxiliary fields
$h_{i}(t)$ in (\ref{eq:Ploc}) have the meaning of effective site
energies, while they are taken here as effective magnetic fields.) In
this way we arrive at 
\begin{equation}\label{eq:dsit1}
\frac{d\langle\sigma_{i}\rangle_{t}}{dt}=-2\langle
w_{i}(\boldsymbol\sigma)\sigma_{i}\rangle_{t}.
\end{equation}
The summation over all $\boldsymbol\sigma$ in the definition of the average on the
right--hand side of (\ref{eq:dsit1}) involves a summation over
$\sigma_{i}=\pm 1$, which we treat with the help of the detailed
balance condition:
\begin{eqnarray}\label{eq:sums}
\sum_{\sigma_{i}}\hspace{-0.2cm}&&\exp[-\beta(H(\boldsymbol\sigma)-h_{i}\sigma_{i})]w_{i}(\boldsymbol\sigma)\sigma_{i} \nonumber\\ 
&&=\frac{1}{2}\sum_{\sigma_{i}}e^{-\beta
  H(\boldsymbol\sigma)}w_{i}(\boldsymbol\sigma)[e^{\beta
  h_{i}\sigma_{i}}\sigma_{i}+e^{-\beta
  h_{i}\sigma_{i}}(-\sigma_{i})]\nonumber \\
&&=\sum_{\sigma_{i}}e^{-\beta
  H(\boldsymbol\sigma)}w_{i}(\boldsymbol\sigma){\rm sinh} \beta h_{i}.
\end{eqnarray}
In the last step we have used ${\rm sinh}(\beta
h_{i}\sigma_{i})=\sigma_{i}\,{\rm sinh}(\beta h_{i})$ and $\sigma_{i}^{2}=1$. To
restore the expression for $P^{(loc)}(\boldsymbol\sigma,t)$ we
multiply and divide (\ref{eq:sums}) by $\exp(\beta h_{i})$. Finally,
it follows from the form of $P^{(loc)}(\boldsymbol\sigma,t)$ that
single--spin averages and the fields ${\bf h}(t)$ are connected by
$h_{i}(t)+h=\partial F/\partial\langle\sigma_{i}\rangle_{t}$, which is
analogous to (\ref{eq:hit}). Here, $F$ is the intrinsic free energy as a
functional of the spin density. Combination of these results with
(\ref{eq:dsit1}) yields (\ref{eq:dsi}) and (\ref{eq:Gamma}).


\begin{thebibliography}{ccccc}
\bibitem{F91} For recent reviews see ``Materials Science and
  Technology Vol. 5: Phase Transformations in Materials'' ed. by
  R.W. Cahn, P. Haasen and E.J. Kramer, (VCH Weinheim, New York, 1991).
\bibitem{Bi74} K. Binder, Z. Physik {\bf 267}, 313 (1974).
\bibitem{Pl97} For a review, see J.F. Gouyet, M. Plapp, W. Dieterich
  and P. Maass, to be published.
\bibitem{Ki79} R. Kikucki, J. Chem. Phys. {\bf 60}, 1071 (1979).
\bibitem{Re94} Several reviews are contained in
  Prog. Theor. Phys. Suppl. {\bf 115} (1994).
\bibitem{Na} M. Nastar, V. Dobretsov and G. Martin, Phil. Mag. A {\bf
  80}, 155 (2000).
\bibitem{Rei96} D. Reinel and W. Dieterich, J. Chem. Phys. {\bf 104},
  5234 (1996).
\bibitem{Fi98} H.P. Fischer, J. Reinhard and W. Dieterich, J.F. Gouyet,
  P. Maass, A. Majhofer and D. Reinel, J. Chem. Phys. {\bf 108}, 3028
  (1998).
\bibitem{Die90} W. Dieterich, H.L. Frisch and A. Majhofer, Z. Physik B
  {\bf 78}, 317 (1990).
\bibitem{Loe93} For a review, see H. L\"{o}wen, Phys. Rep. {\bf 237}, 249 (1993).
\bibitem{Ni93} M. Nieswand, W. Dieterich and A. Majhofer, Phys. Rev. E
  {\bf 47}, 718 (1993);\\
M. Nieswand, A. Majhofer and W. Dieterich, Phys. Rev. E {\bf 48}, 2521
  (1993).
\bibitem{Rei94} D. Reinel, W. Dieterich and A. Majhofer, Phys. Rev. E
  {\bf 50}, 4744 (1994).
\bibitem{KAW73} K. Kawasaki and J.D. Gunton, Phys. Rev. A {\bf 4},
  2048 (1973).
\bibitem{GRA82} H. Grabert, Springer Tracts in Mod. Phys. {\bf 95},
  ed. by G. H\"{o}hler (Springer--Verlag, Berlin, 1982).
\bibitem{LAT00} A. Latz, J. Phys. Cond. Matter {\bf 12}, 6353 (2000).
\bibitem{SCHO94} J. Schofield and I. Oppenheim, Physica A {\bf 204},
  555 (1994).
\bibitem{Ho77} P.C. Hohenberg and B.I. Halperin, Rev. Mod. Phys. {\bf
  49}, 435 (1977).
\bibitem{Per} For a review see J.K. Percus, Acc. Chem. Rev. {\bf 27},
  8 (1994).
\bibitem{Bu1} J. Buschle, P. Maass and W. Dieterich, J. Phys. A:
  Math. Gen. {\bf 33}, L41 (2000).
\bibitem{Bu2} J. Buschle, P. Maass and W. Dieterich, J. Stat. Phys. {\bf
  99}, 273 (2000).
\bibitem{Gla} R.J. Glauber, J. Math. Phys. {\bf 4}, 294 (1963). 
\bibitem{Fus} The Monte Carlo time scale is chosen such that one Monte
  Carlo step per particle corresponds to a time interval $\Delta t$
  equal to the inverse of the largest possible hopping rate in
  (\ref{eq:wik}), $\Delta t=(\alpha\,e^{\beta V})^{-1}$. We take
  $\Delta t$ as our time unit, so that the variable $t$ becomes
  dimensionless. In subsection 4.2, the time unit is $\Delta t=\alpha^{-1}$.
\bibitem{Nas98} R. Nassif, Y. Boughaleb, A. Hekkouri, J.F. Gouyet and
  M. Kolb, Eur. Phys. J. B {\bf 1}, 453 (1998).
\bibitem{Bu3} J. Buschle, P. Maass and W. Dieterich, to be published.
\bibitem{Cr} J. Crank, ``The Mathematics of Diffusion'' (Clarendon
  Press, Oxford, 1975), chapter 10.6.
\end{thebibliography}
\end{document}